\documentclass[12pt,preprint]{aastex}
\usepackage{emulateapj5}
\usepackage{apjfonts}
\usepackage{epsfig}
\usepackage{natbib}

\newcommand{\civ}{\ion{C}{4}}

\newcommand{\etal}{et al.}

\newcommand{\flux}{erg s$^{-1}$ cm$^{-2}$}

\def\gtrsim{\mathrel{\hbox{\rlap{\hbox{\lower4pt\hbox{$\sim$}}}\hbox{\raise2pt\hbox{$>$}}}}}

\newcommand{\hbeta}{H\ensuremath{\beta}}

\newcommand{\lf}{\ensuremath{L_{\rm{5100 \AA}}}}

\newcommand{\lum}{erg s$^{-1}$}

\newcommand{\lledd}{\ensuremath{L_{\mathrm{bol}}/L{\mathrm{_{Edd}}}}}

\newcommand{\mbh}{\ensuremath{M_\mathrm{BH}}}

\newcommand{\mgii}{\ion{Mg}{2}}

\newcommand{\msigma}{\ensuremath{M_{\mathrm{BH}}-\sigmastar}}
\newcommand{\msun}{\ensuremath{M_{\odot}}}

\newcommand{\sigmastar}{\ensuremath{\sigma_{\ast}}}

\def\lax{{$\mathrel{\hbox{\rlap{\hbox{\lower4pt\hbox{$\sim$}}}\hbox{$<$}}}$}}
\def\gax{{$\mathrel{\hbox{\rlap{\hbox{\lower4pt\hbox{$\sim$}}}\hbox{$>$}}}$}}

\slugcomment{To be published in {\it PASP}; Sept. 30, 2009.}
\shorttitle{{\it BH Masses in AGNs}}
\shortauthors{GREENE \& HO}

\begin{document}

\title{Active Galaxies and the Study of Black Hole Demographics}

\author{Jenny E. Greene}
\affil{Department of Astrophysical Sciences, Princeton University, 
Princeton, NJ 08544; Hubble, Princeton-Carnegie Fellow}

\author{Luis C. Ho}
\affil{The Observatories of the Carnegie Institution of Washington, 
813 Santa Barbara Street, Pasadena, CA 91101}

\begin{abstract}

We discuss the critical importance of black hole mass indicators based on 
scaling relations in active galaxies.  We highlight outstanding uncertainties 
in these methods and potential paths to substantial progress in the next 
decade.

\end{abstract}

\section{Black Hole Demographics and Active Galaxies}

The most reliable measures of black hole (BH) mass are derived from
spatially resolved dynamical tracers, the stars at the center of our
own Galaxy providing the most precise BH mass measurement (e.g.,Ghez
et al. 2008; Gillessen \etal\ 2009) followed closely by water masers
in Keplerian rotation fractions of a parsec from the central BH
(e.g.,~Herrnstein \etal\ 2005).  Stellar and gas dynamical models use
the integrated light from stars or gas in the centers of nearby,
relatively inactive galaxies to infer BH masses with $\sim 30\%$
precision (e.g.,~Gebhardt \etal\ 2003), although potential systematics
due to, e.g., the varying importance of dark matter as a function of
galaxy mass are not yet well known (e.g., Gebhardt \& Thomas 2009).
The advent of the \emph{Hubble Space Telescope} has enabled the
measurement of dynamical BH masses in dozens of nearby massive
galaxies, which in turn has revolutionized our understanding of BH
demographics.  It has become clear that not only are supermassive BHs
a ubiquitous component of bulge-dominated galaxy centers (e.g.,~Ho
2004), but that apparently the evolution of the BH and the surrounding
bulge are coupled, as evidenced by the remarkably tight correlations
between BH mass and the properties of the surrounding bulge,
particularly bulge velocity dispersion (the \msigma\ relation;
e.g.,~G{\"u}ltekin \etal\ 2009).

Given the apparent cosmological significance of supermassive BHs in
the evolution of galaxies, we are highly motivated to study the
demographics and growth histories of BHs beyond the Local Universe.
Unfortunately, the dynamical techniques outlined above require very
sensitive observations with high spatial resolution, which are
currently limited to galaxies within tens of Mpc.  Techniques using
active galactic nuclei (AGNs), which currently must be calibrated to
the direct dynamical methods, are the only means to study the
cosmological evolution of BH demographics and accretion.  This article
addresses the current limitations on indirect BH mass measurements,
and experiments that could substantially improve them in the next
decade.

\section{Virial Masses in Active Galaxies}

As early as Woltjer (1959) it was recognized that if the signature
broad emission lines in the spectra of AGNs truly represent gas
orbiting the central BH, then they may serve as a virial tracer of the
enclosed mass, provided one has a scale for the emission region.  A
size can be determined from the delay between variability in the
central continuum source and the surrounding photoionized gas [known
as reverberation mapping (RM); Blandford \& McKee 1982; see review in
Peterson \etal\ 2004].  The BH mass scales as the virial product \mbh
$\propto R~\upsilon^2/G$, but remains uncertain within a scale factor
$f$ due to the unknown structure and kinematics of the broad-line 
region (BLR).  Only with the discovery of the \msigma\
relation did it become possible to independently cross-check
the RM masses, as well as to solve for $f$ in an average sense
(e.g.,~Nelson \etal\ 2004; Onken \etal\ 2004).

RM experiments are time consuming, particularly for luminous objects,
and have been performed for only $\sim 45$ objects to date, but, due
to a well-defined correlation between AGN luminosity and BLR size (the
radius-luminosity relation; e.g.,~Kaspi \etal\ 2000; Bentz \etal\
2009a) it is possible to estimate so-called virial BH masses from
single-epoch AGN spectroscopy.  Comparisons between large samples of
single-epoch virial masses and \sigmastar\ yield reasonable agreement
and similar $f$ values as the RM samples (Greene \& Ho 2006; Shen
\etal\ 2008a).  It is these virial masses that enable study of the
demographics (e.g.,~Greene \etal\ 2008), mass-dependent clustering
(e.g.,~Fine \etal\ 2006; Shen \etal\ 2009), BH mass functions
(e.g.,~Greene \& Ho 2007; Kelly \etal\ 2009b) and Eddington-ratio
distributions (e.g.,~Kollmeier \etal\ 2006; Shen \etal\ 2008b; Hickox
\etal\ 2009) for accreting BHs, as well as investigations of the
evolution in BH-bulge scaling relations with cosmic time (e.g.,~Peng
\etal\ 2006).

While the ability to estimate BH masses using the virial method has
resulted in substantial progress in our understanding of BH
demographics and evolution, the technique is very indirect, and each
calibration step introduces a new layer of uncertainty.  First of all,
the radius-luminosity relation on which the entire architecture rests
has been calibrated using only $\sim 30$ objects.  When measured using
lags between \hbeta\ and \lf\ the slope is found to be $\alpha =
0.519^{+0.063}_{-0.066}$ (Bentz \etal\ 2009a), in agreement with the
\ion{C}{4} slope of $0.52 \pm 0.04$ (Kaspi \etal\ 2007).  However, the
formal errors do not properly reflect the true slope uncertainties.
Prior to the careful galaxy subtraction performed by Bentz et al., the
radius-luminosity relation slope was substantially steeper ($\alpha =
0.67 \pm 0.05$; e.g., Kaspi \etal\ 2005), while the slope also changes
when various objects are removed or included.  In our view the most
significant outstanding systematic uncertainties in the
radius-luminosity relation slope come from the limited scope of the
sample.  The range in luminosity is only \lf$=10^{42} - 10^{46}$~\lum,
with only five and three objects in the lowest and highest decades
respectively, while the Eddington ratios are strongly clustered around
$\sim 10\%$.  There are hardly any radio-loud targets. These biases
are present because only a small sample of local, heterogeneous
targets selected for their known variability properties have been
targeted to date.  Ideally, we would like to see BLR measurements for
BHs ranging over $10^5-10^{9}$~\msun\ and at least two decades in
luminosity at each mass.  {\it Unbiased virial BH masses require a
  measurement of the radius-luminosity relation for objects spanning a
  complete range in BH mass and luminosity.}

Even with a perfect radius-luminosity relation, there is still
ambiguity in the overall mass scaling from object to object due to our
ignorance of BLR structure.  Collin \etal\ (2006) make a compelling
case for systematic changes in the value of $f$ as a function of
accretion rate and perhaps BH mass as well.  Theoretically, we do
expect BLR structure to vary as a function of accretion rate.  For
instance, the existence and opening angle of disk winds are thought to
depend on both \mbh\ and \lledd\ (e.g.,~Proga \& Kallman 2004).
Alternatively, radiation pressure may provide non-negligible support to
the BLR gas, thus biasing the virial masses (Marconi \etal\ 2008)
although debate continues on the importance of this effect in practice
(e.g., Netzer 2009; Onken 2009).  {\it Independent BH mass estimates
  are required to calibrate the f factor for objects spanning a
  complete range in BH mass and luminosity.}

There is one more central concern worth noting here.  Rest-frame
optical spectra containing Balmer emission-lines are not available for
large samples of high-redshift quasars, and we are thus forced to rely
on UV lines, specifically \mgii~$\lambda 2800$~\AA\ and \civ~$\lambda
1550$~\AA\ (e.g., Vestergaard 2002; McLure \& Jarvis 2002).  There is
little reverberation mapping done for \mgii, and thus the virial mass
scale is simply bootstrapped from \hbeta\ (Vestergaard \& Peterson
2006; see also Onken \& Kollmeier 2008).  In the case of \civ,
reverberation mapping has been performed for a small handful of
objects (e.g., Peterson et al. 2005), but the line width may not be
dominated by virial motions. It is clear that the scatter between
\civ\ and \mgii\ or \hbeta\ line widths is substantial (e.g., Baskin
\& Laor 2005; Shen et al. 2008), and the overall reliability of \civ\
as a virial estimator remains a matter of debate (e.g., Kelly \etal\
2007).  We choose here to focus on improving the Balmer line
calibrations, as a necessary step to understanding the UV lines.

\section{A Wish List}

We focus on a few potential avenues for future progress in improving the 
reliability of BH mass measurements that we believe are achievable with 
existing or planned instrumentation:

\begin{enumerate}

\item 
Dedicated RM campaigns are required to securely measure the
radius-luminosity relation for a large and representative sample of
active galaxies.  A recent extensive monitoring program at Lick
Observatory demonstrates the feasibility of such campaigns (Bentz et
al.  2009b; Denney et al. 2009), measuring an additional eight new
BLR radii in a low-luminosity regime as yet unexplored.  The data
are of such high quality that velocity-resolved lags can be measured
in a few cases, which may yield measurements of $f$ for individual
objects (Bentz \etal\ 2008, 2009b; Denney et al. 2009).

\item
A dedicated spectroscopic monitoring telescope could substantially
improve the diversity of the samples used to define the
radius-luminosity relation at intermediate luminosities and low
redshift.  It is convenient to consider three distinct luminosity
regimes, 

\hskip -0.4in
\psfig{file=rm_other.epsi,width=0.5\textwidth,keepaspectratio=true}
\vskip -0mm
\figcaption[]{
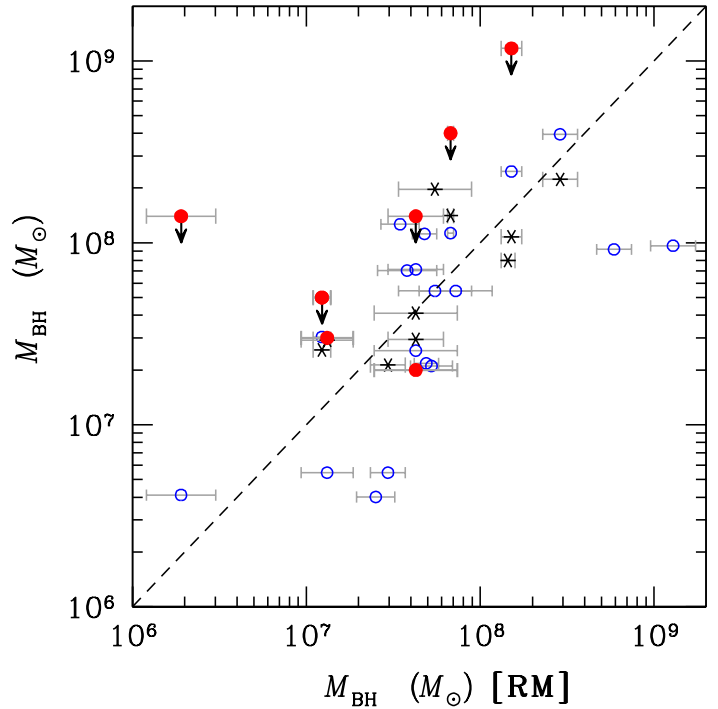
Comparison of the masses derived using RM and
various other techniques.  For simplicity, we adopt the RM sample
and mass estimates of Bentz \etal\ (2009a). Gas-dynamical mass
estimates and upper limits ({\it filled red circles}) are from Hicks
\& Malkan (2009), masses inferred from stellar velocity dispersions
({\it open blue circles}) are from Nelson \etal\ (2004), Onken
\etal\ (2004), and Dasyra \etal\ (2007), and masses from X-ray
variability ({\it stars}) are from Niko{\l}ajuk \etal\ (2006).  The
one-to-one relation is shown only to guide the eye ({\it dashed
  line}). Note that the absolute RM mass scale ($f$) has been set to
match a subset of the blue points.  NGC 4395, with a RM mass of 
$\sim 10^5$~\msun\ (Peterson \etal\ 2005) and 
a $\sigma_{\ast}$-based mass of $\sim 4 \times 10^4$~\msun\ 
(Filippenko \& Ho 2003) does not appear 
in the figure only because of the very low mass.
\label{fits}}
\vskip 5mm
\noindent
and, for this discussion, luminosity refers to \lf.  In
general, campaigns should last at least 3 times as long as the
maximum timescale that they need to probe (e.g.,~Peterson 2001),
although recent monitoring campaigns demonstrate that $\sim 6$ times
longer is actually preferred in the presence of finite sampling and
signal-to-noise ratio (e.g.,~Bentz \etal\ 2008).  Of course, only a
fraction of sources will vary at any given time.  Objects with
\lf$\lesssim 10^{42}$~\lum\ have lags less than one light day and
require continuous monitoring with an 8 m-class telescope over a
period of at least a week, in order to achieve sufficiently short
individual exposures to resolve the lag times but still detect
broad-line variability.  So far, the only such object with
successful RM in this regime is NGC 4395, which was done with \civ\
$\lambda$1549 from space (Peterson \etal\ 2005).  For the regime
\lf$= 10^{42}-10^{43}$~\lum, the lags are a few days and the targets
should be observed at least every other day for a few months
(e.g.,~Bentz \etal\ 2008).  At \lf$=10^{43}-10^{45}$~\lum\ the
appropriate sampling is $\sim$ weekly for 1--2 years.  Objects more
luminous require significantly longer baselines, since not only are
their BLRs physically larger, but they also tend to reside at $z >
0.5$, causing additional stretching of the observed variability
timescale.

The targets requiring daily observations must be monitored
photometrically with a dedicated 1 m-class telescope.  For more
luminous targets the 3--5 day temporal sampling delivered by current
and upcoming time-domain surveys (e.g., Palomar Transient Factory, Law
\etal\ 2009; 
Pan-STARRS\footnote{http://pan-starrs.ifa.hawaii.edu/public/}; LSST,
Ivezic \etal\ 2008) should provide adequate photometry.  Assuming a 4
m-class spectroscopic telescope, with a 2\arcsec\ slit (to minimize
light losses in bad seeing conditions) and $\sim 10$~\AA\ resolution,
it should be possible to obtain sufficient S/N in $\lesssim 1$ hr to
measure $\sim 10 \%$ variability in the \hbeta\ line for objects with
$F_{\rm \hbeta} \gtrsim 10^{-14}$~\flux\ (e.g.,~Kaspi \etal\ 2000).
We note that many recent lag measurements have achieved $1 \%$
spectrophotometry in the emission lines (e.g., Denney \etal\ 2006;
Bentz \etal\ 2006, 2008), but such precision is feasible only for a
very small number of targets.  In the Sloan Digital Sky Survey Data
Release 4 (Adelman-McCarthy \etal\ 2006) at $z \lesssim 0.05$ there
are $\sim 100$ broad-line objects with \lf$=2\times 10^{42} - 4 \times
10^{43}$~\lum\ and \mbh\ $\approx 10^5-10^8$ \msun\ (Greene \& Ho 2007).
Assuming that 1 hr of integration is spent per observation and that
each galaxy needs $\sim 90$ observations to determine a lag, a
two-year dedicated spectroscopic survey could observe more than half
of the entire sample, if only one-fifth of the observing time is lost
to bad weather.  This very crude sketch demonstrates that even with
existing modest resources, substantial progress could be made on
measurements of the radius-luminosity relation as a function of AGN
properties.  To significantly expand the dynamic range in
luminosities, however, requires both an investment of 8 m-class
resources for low-luminosity targets and a much longer campaign to
target luminous (and necessarily distant) ones (e.g.,~Kaspi \etal\
2007).

\item
In addition to improvements in the radius-luminosity relation, independent
measurements of BH mass are needed, again spanning the full
mass-luminosity parameter space, with which to calibrate $f$.  To
zeroth order $f$ has been calculated by scaling RM or virial masses to
match the \msigma\ relation (see Fig. 1).  In a handful of cases it ought to be
possible to constrain the kinematic structure of the BLR using
two-dimensional RM in which lags are determined as a function of
velocity in the broad line (e.g.,~Bentz et al. 2008).  While
expensive, this technique provides a real measurement of BLR
structure.

The next obvious step is to amass a sufficiently large sample of
stellar velocity dispersions in AGNs, spanning a wide enough range in
BH mass, that $f$ can be measured as a function of linewidth and
luminosity. Some progress has been made using both near-infrared
spectroscopy (e.g.,~Dasyra \etal\ 2007) and adaptive-optics assisted
integral-field unit spectroscopy (e.g,~Watson \etal\ 2008).  With
large samples (e.g.,~10 objects per decade in mass and luminosity) it
should be possible to measure $f$ as a function of \mbh\ and $L$.

Comparison with \msigma\ in the determination of $f$ may well fail if
the galaxies in question do not obey the \msigma\ relation.  It is
important to bear in mind that while the range of dynamical BH masses
extends over \mbh\ $\approx 10^6-10^8$~\msun, the majority of the
objects are clustered around \mbh\ $\approx 10^8$~\msun.  In contrast,
the masses of the local active samples range from $\sim 10^5$ to
$10^9$~\msun, and the majority of those with \sigmastar\ measurements
have \mbh\ $\approx 10^7$~\msun.  There is some indication that the
scatter, and possibly the shape, of the \msigma\ relation depend on
galaxy type (e.g.,~Hu 2008; Greene \etal\ 2008; C. Y. Peng \&
L. C. Ho, in preparation; Greene, J. E. \etal, in preparation).
Furthermore, active systems may not obey the \msigma\ relation during
their active phase (e.g.,~Ho \etal\ 2008; Kim \etal\ 2008).  Direct
gas or stellar dynamical BH masses are needed to bypass these
degeneracies, but so far have been attempted in only a handful of
cases (Fig. 1; Davies \etal\ 2006; Onken \etal\ 2007; Hicks \& Malkan
2008).  Of the RM objects with stellar velocity dispersions in hand,
nine are close enough to attempt a dynamical measurement
(i.e.,~current adaptive-optics assisted spectrographs can reach to a
few times the gravitational sphere of influence of the BH; e.g.,~Hicks
\& Malkan 2008).  The number should increase with new RM surveys
targeting low-luminosity sources (e.g.,~Bentz et al. 2008).  While
these numbers are still small, and restricted to $\sim 10^7$~\msun\
BHs, even a few dynamical BH mass measurements would provide a crucial
cross-check on RM.

\item 
Aside from direct dynamical masses and galaxy scaling relations,
variability timescales also correlate with \mbh.  In particular, it
has been shown that the characteristic X-ray variability timescale
correlates with BH mass (e.g.,~Czerny et al. 2001; Lu \& Yu 2001;
Markowitz et al. 2003; ; Niko{\l}ajuk et al. 2004, 2006; Done \&
Gierli{\'n}ski 2005).  In principle, X-ray variability is completely
independent of the \msigma\ relation and RM, since the masses are
scaled to the variability properties of high-mass X-ray binaries.
However, the story may be more complicated, as there appears to be
additional dependence on the luminosity or Eddington ratio of the BH
(McHardy et al. 2006).  At the moment, X-ray power-spectrum analysis
has been possible only for a handful of bright local AGNs
(e.g.,~Uttley \etal\ 2002).  Upcoming all-sky X-ray monitors such as
\emph{Lobster}\footnote{{\tt www.mpe.mpg.de/erosita/MDD-6.pdf}}
(0.1--3 keV) and \emph{MAXI} (2--30 keV; Matsuoka \etal\ 2009) will
deliver increases of factors of a few in the number of targets with
adequate long-term monitoring to explore all scales in X-ray
variability. For instance, \emph{Lobster}, with a sensitivity of
0.15 mCrab per day, will monitor $\sim 400$ AGNs per year with daily
cadence and $\sim 20\%$ accuracy, and ten percent of those will have
$\sim 5\%$ accuracy. Thus, the opportunities for cross-calibration
between X-ray variability, dynamical masses, RM, and virial mass
estimates will also increase dramatically.

\item
Much like in the X-rays, optical variability timescales are correlated
with BH mass, again with a secondary dependence on luminosity
(e.g.,~Collier \& Peterson 2001; Kelly et al. 2009a; Walsh et
al. 2009).  Again, upcoming optical time-domain surveys (see above)
will enable cross-calibration of this relatively new technique with
large and relatively unbiased samples.  We note also that a byproduct
of a large RM campaign would be very well-sampled optical light curves
yielding characteristic optical timescales that can be compared with
the RM masses.

\end{enumerate}

In closing, we note that our discussion focuses on observational
avenues for improvements in BH mass measurement techniques.  While we
believe the experiments described above are crucial to tie down the BH
mass scale at cosmological distances, they alone are not
sufficient.  Despite decades of research, we still do not have a
complete model for the BLR.  We are hopeful that these experiments,
in addition to providing more robust empirical estimators of BH mass,
will also provide needed constraints on the kinematics and structure
of the line-emitting region as a function of \mbh, thereby inspiring more 
detailed modeling of the physics of the BLR.

\acknowledgements

We thank M.~Bentz for helpful conversations, and we thank the referee, 
B.~M.~Peterson for a thorough and thought-provoking report.


\begin{thebibliography}{}

\bibitem[]{}Adelman-McCarthy, J.~K., et al.\ 2006, \apjs, 162, 38 

\bibitem[]{}Baskin, A., \& Laor, A. 2005, \mnras, 356, 1029 

\bibitem[]{}Bentz, M.~C., et al. 2006, \apj, 651, 775 

\bibitem[]{}------. 2008, \apj, 689, L21 

\bibitem[]{}Bentz, M.~C., Peterson, B.~M., Netzer, H., Pogge, R.~W., 
\& Vestergaard, M. 2009a, \apj, 697, 160 

\bibitem[]{}Bentz, M.~C., et al. 2009, \apj, submitted (astroph:0908.0003)

\bibitem[]{}Blandford, R.~D., \& McKee, C.~F. 1982, \apj, 255, 419 

\bibitem[]{}Collier, S., \& Peterson, B.~M. 2001, \apj, 555, 775 

\bibitem[]{}Collin, S., Kawaguchi, T., Peterson, B.~M., \& Vestergaard, M. 
2006, \aap, 456, 75 

\bibitem[]{}Czerny, B., Niko{\l}ajuk, M., Piasecki, M., 
\& Kuraszkiewicz, J. 2001, \mnras, 325, 865 

\bibitem[]{}Dasyra, K.~M., et al. 2007, \apj, 657, 102 

\bibitem[]{}Davies, R.~I., et al. 2006, \apj, 646, 754 

\bibitem[]{}Denney, K.~D., et al. 2006, \apj, 653, 152 

\bibitem[]{}------. 2009, \apj, submitted (astroph/0908.0327) 

\bibitem[]{}Done, C., \& Gierli{\'n}ski, M. 2005, \mnras, 364, 208 

\bibitem[]{}Filippenko, A.~V., \& Ho, L.~C. 2003, \apj, 588, L13

\bibitem[]{}Fine, S., et al. 2006, \mnras, 373, 613 

\bibitem[]{}Gebhardt, K., \& Thomas, J. 2009, \apj, 700, 1690 

\bibitem[]{}Gebhardt, K., et al. 2003, \apj, 583, 92 

\bibitem[]{}Ghez, A.~M., et al. 2008, \apj, 689, 1044 

\bibitem[]{}Gillessen, S., Eisenhauer, F., Trippe, S., 
Alexander, T., Genzel, R., Martins, F., \& Ott, T. 2009, \apj, 692, 1075 

\bibitem[]{}Greene, J.~E., \& Ho, L.~C. 2006, \apj, 641, L21

\bibitem[]{}------. 2007, \apj, 667, 131 

\bibitem[]{}Greene, J.~E., Ho, L.~C., \& Barth, A.~J. 2008, \apj, 688, 159 

\bibitem[]{}G{\"u}ltekin, K., et al. 2009, \apj, 698, 198 

\bibitem[]{}Herrnstein, J.~R., Moran, J.~M., Greenhill, L.~J., 
\& Trotter, A.~S. 2005, \apj, 629, 719 

\bibitem[]{}Hickox, R.~C., et al. 2009, \apj, 696, 891 

\bibitem[]{}Hicks, E.~K.~S., \& Malkan, M.~A. 2008, \apjs, 174, 31 

\bibitem[]{}Ho, L. C., 2004, ed., Carnegie Observatories Astrophysics Series, 
Vol. 1: Coevolution of Black Holes and Galaxies 
(Cambridge: Cambridge Univ. Press)

\bibitem[]{}Ho, L.~C., Darling, J., \& Greene, J.~E. 2008, \apj, 681, 128 

\bibitem[]{}Hu, J. 2008, \mnras, 386, 2242 

\bibitem[]{}Ivezic, Z., Tyson, 
J.~A., Allsman, R., Andrew, J., Angel, R., 
\& for the LSST Collaboration 2008, (astroph/0805.2366)

\bibitem[]{}Kaspi, S., Brandt, W.~N., Maoz, D., Netzer, H., Schneider, D.~P., \&
Shemmer, O. 2007, \apj, 659, 997

\bibitem[]{}Kaspi, S., Maoz, D., Netzer, H., Peterson, B.~M., Vestergaard, M., 
\& Jannuzi, B.~T. 2005, \apj, 629, 61 

\bibitem[]{}Kaspi, S., Smith, P.~S., Netzer, H., Maoz, D., 
Jannuzi, B.~T., \& Giveon, U. 2000, \apj, 533, 631 

\bibitem[]{}Kelly, B.~C., \& Bechtold, J. 2007, \apjs, 168, 1 

\bibitem[]{}Kelly, B.~C., Bechtold, J., \& Siemiginowska, A. 
2009a, \apj, 698, 895 

\bibitem[]{}Kelly, B.~C., Vestergaard, M., \& Fan, X. 2009b, \apj, 692, 1388 

\bibitem[]{}Kim, M., Ho, L.~C., Peng, 
C.~Y., Barth, A.~J., Im, M., Martini, P., 
\& Nelson, C.~H. 2008, \apj, 687, 767 

\bibitem[]{}Kollmeier, J.~A., et al. 2006, \apj, 648, 128 

\bibitem[]{}Law, N.~M., et al. 2009, \pasp, submitted (astroph/0906.5350)

\bibitem[]{}Lu, Y., \& Yu, Q. 2001, \mnras, 324, 653 

\bibitem[]{}Marconi, A., Axon, 
D.~J., Maiolino, R., Nagao, T., Pastorini, G., Pietrini, P., Robinson, A., 
\& Torricelli, G. 2008, \apj, 678, 693 

\bibitem[]{}Markowitz, A., et al. 2003, \apj, 593, 96 

\bibitem[]{}Matsuoka, M., et al. 2009, \pasj, accepted (astroph/0906.0631)

\bibitem[]{}McHardy, I.~M., Koerding, E., Knigge, C., Uttley, P., 
\& Fender, R.~P. 2006, \nat, 444, 730 

\bibitem[]{}McLure, R.~J., \& Jarvis, M.~J. 2002, \mnras, 337, 109

\bibitem[]{}Nelson, C.~H., Green, 
R.~F., Bower, G., Gebhardt, K., \& Weistrop, D. 2004, \apj, 615, 652 

\bibitem[]{}Netzer, H. 2009, \apj, 695, 793
 
\bibitem[]{}Niko{\l}ajuk, M., Czerny, B., Zi{\'o}{\l}kowski, J., 
\& Gierli{\'n}ski, M. 2006, \mnras, 370, 1534 

\bibitem[]{}Nikolajuk, M., Papadakis, I.~E., \& Czerny, B. 
2004, \mnras, 350, L26 

\bibitem[]{}Onken, C.~A. 2009, \apj, submitted (astroph/0907.4192)

\bibitem[]{}Onken, C.~A., et al. 2007, \apj, 670, 105 

\bibitem[]{}Onken, C.~A., Ferrarese, 
L., Merritt, D., Peterson, B.~M., Pogge, R.~W., Vestergaard, M., 
\& Wandel, A. 2004, \apj, 615, 645 

\bibitem[]{}Onken, C.~A., \& Kollmeier, J.~A. 2008, \apj, 689, L13 

\bibitem[]{}Peng, C.~Y., Impey, C.~D., 
Rix, H.-W., Kochanek, C.~S., Keeton, C.~R., Falco, E.~E., Leh{\'a}r, J., 
\& McLeod, B.~A. 2006, \apj, 649, 616 

\bibitem[]{}Peterson, B.~M. 2001, in ``The Starburst-AGN Connection 2001'',
 eds. I. Aretxaga, D. Kunth, \& R. M{\'u}jica (Singapore, World Scientific), 3

\bibitem[]{}Peterson, B.~M., et al. 2004, \apj, 613, 682 

\bibitem[]{}------. 2005, \apj, 632, 799 

\bibitem[]{}Proga, D., \& Kallman, T.~R. 2004, \apj, 616, 688 

\bibitem[]{}Shen, J., Vanden Berk, D.~E., Schneider, D.~P., 
\& Hall, P.~B. 2008a, \aj, 135, 928 

\bibitem[]{}Shen, Y., et al. 2009, \apj, 697, 1656 

\bibitem[]{}Shen, Y., Greene, J.~E., Strauss, M.~A., Richards, G.~T., 
\& Schneider, D.~P. 2008b, \apj, 680, 169 

\bibitem[]{}Uttley, P., McHardy, I.~M., \& Papdakis, I.~E. 2002, 
\mnras, 332, 231

\bibitem[]{}Vestergaard, M. 2002, \apj, 571, 733 

\bibitem[]{}Vestergaard, M., \& Peterson, B.~M. 2006, \apj, 641, 689 

\bibitem[]{}Walsh, J.~L., \etal\ 2009, \apj, submitted

\bibitem[]{}Watson, L.~C., Martini, P., Dasyra, K.~M., 
Bentz, M.~C., Ferrarese, L., Peterson, B.~M., Pogge, 
R.~W., \& Tacconi, L.~J. 2008, \apjl, 682, L21 

\bibitem[]{}Woltjer, L. 1959, \apj, 130, 38 

\end{thebibliography}
\end{document}